\documentclass[12pt]{article}
\usepackage{graphicx}
\usepackage{hyperref}
\usepackage{subfigure}
\usepackage{epsfig}
\usepackage{amsmath}
\usepackage{amsfonts}
\usepackage{amssymb}

\def\[{\left [}
\def\]{\right ]}
\def\({\left (}
\def\){\right )}
\begin{document}
\pagestyle{empty}

\vspace*{5mm}

\begin{center}
{\LARGE \bf  Fermion Flavor in Soft-Wall AdS}

\vspace{1.0cm}

{\sc Tony Gherghetta$^{a,}$}\footnote{E-mail:  tgher@unimelb.edu.au}
{\small and}
{\sc Daniel Sword$^{b,}$}\footnote{E-mail:  sword@physics.umn.edu}
\\
\vspace{.5cm}
{\it\small {$^a$School of Physics, University of Melbourne, Victoria 3010, Australia}}\\
{\it\small {$^b$School of Physics and Astronomy, University of Minnesota,\\
Minneapolis, Minnesota 55455, USA}}
\end{center}


\vspace{1cm}

\begin{abstract}
The formalism for modeling multiple fermion generations in a warped extra dimension with a soft-wall is presented. A bulk Higgs condensate is responsible for generating mass for the zero-mode fermions but leads to additional complexity from large mixing between different flavors. We extend existing single-generation analyses by considering new special cases in which analytical solutions can be derived and discuss flavor constraints. The general three-generation case is then treated using a simple numerical routine. Assuming anarchic 5D parameters we find a fermion mass spectrum resembling the standard model quarks and leptons with highly degenerate couplings to Kaluza-Klein gauge bosons. This confirms that the soft-wall model has similar attractive features as that found in hard-wall models, providing a framework to generalize existing phenomenological analyses.
\end{abstract}

\newpage
\setcounter{page}{1}
\setcounter{footnote}{0}
\pagestyle{plain}

\section{Introduction}\label{intro}

The warped extra dimension provides an alternative framework in which to address the Standard Model (SM) gauge hierarchy problem~\cite{Randall:1999ee} and the fermion mass hierarchy~\cite{Grossman:1999ra, Gherghetta:2000qt}. It is a particularly attractive scenario because, by the AdS/CFT correspondence~\cite{Maldacena:1997re}, the five-dimensional (5D) framework is dual to a four-dimensional (4D) strongly-coupled conformal field theory. This allows the physics of the warped fifth dimension to be recast in terms of 4D strong dynamics. In particular, an infrared (IR) brane used to generate the Higgs cutoff scale is interpreted in the 4D dual as the breaking of conformal symmetry at low energy and the generation of a mass-gap by an operator of infinite scaling dimension. While the essential physics is captured in this theoretically idealized situation, it is more natural to expect operators of finite dimension in the dual theory. This can be achieved in soft-wall models in which the ``hard-wall" IR brane is replaced by a scalar field (the ``dilaton") whose nontrivial bulk profile corresponds to conformal symmetry breaking in the dual 4D theory.

The Standard Model in the soft-wall warped dimension was considered in Ref.\cite{Batell:2008me}. Since there is no IR brane, SM fields are necessarily bulk fields, which includes not only the gauge bosons and fermions, but also the Higgs field. Even though the fifth-dimension is semi-infinite, the dilaton does provide a dynamical cutoff to the warped dimension. This leads to a discrete Kaluza-Klein (KK) mass spectrum, but with the distinctive feature that there is a variety of KK spacing between the resonances, including linear Regge-like behavior as in QCD. A bulk Higgs condensate is responsible for breaking electroweak symmetry but causes the analysis of fermions to be particularly involved. An analytical solution can nevertheless be obtained in the case of a single fermion generation confirming that the nice features of hard-wall scenarios, such as fermion mass hierarchies and universal KK gauge couplings, also exist with the soft wall~\cite{Gherghetta:2000qt}. However these features have yet to be confirmed in a complete three-generation soft-wall model. 

In this paper we present a more comprehensive analysis of bulk fermions in a soft-wall warped dimension (see also \cite{Delgado:2009xb,Aybat:2009mk}). While new analytical solutions are found for special cases in the case of a single generation, the complete three-generation case can only be treated numerically. This is because the bulk Higgs condensate causes large mixing between fermion flavors in the equations of motion which makes finding analytical solutions nontrivial. Nonetheless, numerical techniques can be used and we present a numerical routine that can solve the general problem with arbitrary 5D mass parameters. Importantly we find that starting with ``anarchic" 5D parameters we are able to generate fermion mass hierarchies and universal couplings to KK gauge bosons, analogous to that found in hard-wall models. This provides a framework to generalize existing phenomenological analyses to include the soft-wall Standard Model.

The layout of this paper is as follows. In Section \ref{setup}, we review the setup needed to model fermions in the soft-wall background and present the fermion equations of motion. In Section \ref{spectrum}, we develop the tools needed to solve the equations of motion in the special cases where the equations can be partially decoupled. One of these cases has been detailed before in Refs. \cite{Batell:2008me, Delgado:2009xb}, while the remaining cases are new. We then show that these additional analytical cases can be used to reproduce many of the recent numerical results of \cite{Aybat:2009mk}. We conclude this section by discussing flavor changing neutral current processes and by detailing the couplings of fermions to gauge bosons--an analysis which is easily generalized to other bulk couplings. In Section \ref{numerical}, we present a very simple, non-iterative routine which can be used to analyze multiple generations of fermions in an arbitrary background. We present the full dependence of SM fermion masses on the 5D bulk mass parameters and compare the results to a typical hard-wall model. The behavior is shown to be very different in the phenomenologically interesting region of the parameter space, where the bulk $SU(2)_L$ doublet and singlet fermions have opposite bulk masses. We then present results for the case of three fermion generations with substantial mixing between bulk profiles, and find example spectra resembling the up- and down-type quarks (and charged leptons) in the spirit of Ref.\cite{Grossman:2004rm}.

\section{Fermions in the Soft-Wall Background}\label{setup}

We work in a 5D spacetime $(x^\mu,z)$ with conformal coordinate $z$ and metric:
\begin{equation}
ds^2 = e^{-2 A(z)} \eta_{MN}dx^M dx^N~,
\label{swmetric}
\end{equation}
where $\eta_{MN}={\rm diag}(-,+,+,+,+)$. In particular we will consider a pure AdS metric, i.e. $A(z)=\log kz$ with 
$k$ the AdS curvature scale. The spacetime is defined on the interval $z\in [z_0,\infty)$, where $z_0$ is the 
location of the ultraviolet (UV) brane. Though the spacetime extends to $z\to\infty$, we have in mind a soft-wall setup in which 
the dilaton, $\Phi$ obtains a background value and provides a dynamical cutoff to spacetime along the 
fifth coordinate $z$. In this scenario, gauge and matter fields are described by the action,
\begin{equation}
S=\int d^5 x \sqrt{-g}\, e^{-\Phi} {\cal L}, 
\label{matteraction}
\end{equation}
where $\cal L$ is the 5D Lagrangian. While much of our discussion of fermions is valid in general, for the sake 
of concreteness we will specifically consider a dilaton profile given by:
\begin{equation}\label{dilaton}
\Phi(z) = (\mu z)^2,
\end{equation}
where the soft-wall mass scale $\mu\sim1~{\rm TeV}$. This form for the dilaton was also studied in detail in Ref.\cite{Batell:2008me}. 

Consider 5D Dirac fermions, $\Psi_L$ ($\Psi_R$) which transform as a doublet (singlet) under $SU(2)_L$. 
It is straightforward to embed our setup in a theory with a bulk custodial $SU(2)_L\times SU(2)_R$ symmetry, but this will not be essential for our discussion.
In the absence of Yukawa interactions, the fermion action is given by:
\begin{eqnarray}
S&=&-\int d^5x \sqrt{-g}~e^{-\Phi}\left[~ \frac{1}{2}\left( {\bar\Psi}^{a i}_L e^M_A \gamma^A D_M \Psi^{a i}_L - D_M {\bar\Psi}^{a i}_L e^M_A \gamma^A \Psi^{a i}_L\right) + M_L^{ij}{\bar\Psi}^{a i}_L \Psi^{a j}_L\right.  \nonumber \\
&&\qquad\qquad+\left. \frac{1}{2}\left( {\bar\Psi}^{i}_R e^M_A \gamma^A D_M \Psi^{i}_R - D_M {\bar\Psi}^{i}_R e^M_A \gamma^A \Psi^{i}_R\right) + M_R^{ij}{\bar\Psi}^{i}_R \Psi^{j}_R \right]~,
\label{ferma}
\end{eqnarray}
where $e^M_A=e^{A}\delta^M_A$ is the vielbein and $D_M=\partial_M +\omega_M$ is the covariant derivative with spin connection $\omega_M$. The index $a$ is an $SU(2)$ label, while $i,j$ are 5D flavor indices. 

The projections of the Dirac spinors are given by $\Psi^{a i}_{L\pm}=\pm \gamma^5 \Psi^{a i}_{L\pm}$ and similarly for $\Psi^{i}_R$. Dirichlet conditions are imposed on the fields $\Psi_{L-}^{i a}$ and $\Psi_{R+}^{i}$ at the UV boundary:
\begin{eqnarray}
\Psi^{a i}_{L-}(x,z) \big\vert_{z_0} &=&0,  \nonumber \\
\Psi^{i}_{R+}(x,z) \big\vert_{z_0} &=&0.
\label{bcdir}
\end{eqnarray}
Without bulk Yukawa interactions these boundary conditions give rise to massless chiral fermions from the 4D point of view. These zero-modes can obtain a mass by introducing a Yukawa coupling to the Higgs, whose vacuum expectation value (VEV) is $z$-dependent. The Yukawa interaction contribution to the action is:
\begin{eqnarray}
S_{Yukawa}&=&-\int d^5 x \sqrt{-g} e^{-\Phi} \[\,\frac{\lambda^{ij}_5}{\sqrt{k}}\,{\bar\Psi}^{a i}_L(x,z) H^{a}(x,z)\Psi^{j}_R(x,z) +{\rm h.c.}\, \], \nonumber\\
&\equiv& -\int d^5 x \sqrt{-g} e^{-\Phi} \, \Big[ m^{ij}(z)\,{\bar\Psi}^i_L(x,z)\Psi^j_R(x,z) +{\rm h.c.}\, \Big],
\end{eqnarray}
where we have substituted the background value for the Higgs field:
\begin{equation}
H(x,z)\rightarrow H(z)=\frac{h(z)}{\sqrt{2}}\left(\begin{array}{c} 0\\1 \end{array}\right),
\end{equation}
and dropped the $SU(2)$ labels $\Psi_L\equiv\Psi_L^2$. The effective $z$-dependent bulk mass term arising from the Yukawa interaction is simply:
\begin{equation}
m^{ij}(z)\equiv \frac{\lambda^{ij}_5}{\sqrt{2\,k}}h(z).
\end{equation}
To ensure a discrete spectrum of fermion masses, the Higgs VEV must grow faster than the metric 
factor, $e^{A(z)}=1/(k z)$, decays. Namely,\footnote{Other possiblities may also be considered. For example, 
if $\lim\limits_{z\to\infty} h(z)/z\to\mu > 0$, there can exist discrete low-lying modes with a continuous spectrum 
above a ``mass gap", as in Refs. \cite{Falkowski:2008yr,Delgado:2009xb}.}
\begin{equation}
\lim\limits_{z\to\infty} \frac{h(z)}{z}\to\infty.
\end{equation}
Varying the action with respect to ${\bar\Psi}_{L,R}$, we find the equations of motion: 
\begin{eqnarray} 
\gamma^\mu \partial_\mu \psi^i_{L\pm} 
\mp \partial_5\psi^i_{L\mp}
 + e^{-A}M_L^{ij}\psi^j_{L\mp}
  + e^{-A}m^{ij}\psi^j_{R\mp} & = & 0, \\
  \gamma^\mu \partial_\mu \psi^i_{R\pm} 
\mp \partial_5\psi^i_{R\mp}
 + e^{-A}M_R^{ij}\psi^j_{R\mp}
  + e^{-A}m^{\dag ij} \psi^j_{L\mp} & = & 0,
\end{eqnarray}
where we have defined $\Psi=e^{2A+\Phi/2}\psi$. This transformation shows that the fermion mass spectra 
do not depend on the presence of the dilaton. Rather, it is the Higgs VEV that sets the fermion spacing, 
in contrast to the case of bosonic fields.

The KK expansion for the fields $\psi_{L,R\pm}$ is assumed to be:
\begin{eqnarray}
\psi^i_{L\pm}(x,z)&=&\sum_{n,\alpha} f^{i \alpha(n)}_{L\pm}(z)\psi^{\alpha (n)}_{\pm}(x), \label{expf1} \\
\psi^i_{R\pm}(x,z)&=&\sum_{n,\alpha} f^{i \alpha(n)}_{R\pm}(z)\psi^{\alpha (n)}_{\pm}(x),
\label{expf2}
\end{eqnarray}
where $\gamma^{\mu}\partial_{\mu}\psi_{\pm}^{\alpha (n)} = -m_n^{\alpha}\psi_{\mp}^{\alpha (n)}$ (no sum over $\alpha$). Similar to the conventions of Ref.\cite{Grossman:2004rm} we have introduced separate Latin and Greek indices labelling the 5D and 4D flavor, respectively. Defining the vectors:
\begin{equation}
f^{i \alpha (n)}_{\pm}  =
\left(\begin{array}{c} 
 f^{i \alpha (n)}_{L\pm} \\
 f^{i \alpha (n)}_{R\pm}
\end{array} \right), 
\label{fvec}
\end{equation}
allows the equations of motion for the 5D fields to be written in the form:
\begin{equation}
\left[ \pm \partial_5  \delta^{ij}  + {\cal M}^{ij}\right] f^{j \alpha (n)}_{\pm}(z) = m^{\alpha}_n f^{i \alpha (n)}_\mp,
\label{genferm}
\end{equation}
where the mixing matrix is defined as
\begin{equation}
{\cal M}=
e^{-A}\left( 
\begin{array}{cc} 
M^{ij}_L  & m^{ij}(z) \\
m^{\dag ij}(z)& M^{ij}_R 
\end{array} \right). 
\label{matrix} 
\end{equation}
Note that $\alpha, i, j$ run from $1,\ldots N_F$, where $N_F$ is the number of fermion generations. Thus ${\cal M}^{ij}$ is a $2N_F\times 2N_F$ matrix, and Eq.\eqref{genferm} represents a coupled system of $4N_F$ differential equations for each $\alpha$. The 4D fermion fields $\psi_\pm^{\alpha (n)}(x)$ are canonically normalized by requiring that:
\begin{equation}
\int_{z_0}^\infty dz\,\[(f^{i \alpha (n)}_{L\pm})^{\dag} f^{i \beta (m)}_{L\pm}+(f^{i \alpha (n)}_{R\pm})^{\dag}f^{i \beta (m)}_{R\pm} \]=\delta^{nm}\delta^{\alpha \beta}.
\label{normf}
\end{equation}
Note that the index $i$ is to be summed over in this expression.
This completes the general discussion of the fermion setup. To obtain the spectrum of fermion masses, 
the equations of motion \eqref{genferm} are solved subject to the boundary conditions \eqref{bcdir} and orthonormality conditions \eqref{normf}.

\section{Fermion Spectrum}\label{spectrum}

The coupled equations \eqref{genferm} cannot be solved analytically except for a few special cases, depending upon the particular form of the Higgs VEV and the relative bulk masses for the fermions. Up to this point, the only solvable cases that have been presented in the literature have involved just a single generation with degenerate bulk masses for the fields $\Psi_L$ and $\Psi_R$. As it turns out, in AdS space there are additional special cases which allow for the second-order equations of motion to be diagonalized and solved exactly. The collection of solvable cases provides a qualitatively complete picture of fermion behavior in the entire parameter space.

Next, we review the single generation case in detail. We begin very generally, emphasizing that these methods apply to a wide variety of soft-wall models in AdS. We then specialize to a quadratic VEV and solve the equations of motion directly for the special cases. The analytic solutions allow us to verify the results of the numerical routine we 
present in Section \ref{numerical} (as well as a recent numerical treatment in which the Yukawas are treated perturbatively~\cite{Aybat:2009mk}).

\subsection{Single Generation}\label{singlegensection}
For a single generation of fermions, equation \eqref{genferm} becomes:
\begin{equation}\label{singlegenferm}
\(\pm\partial_z+{\cal M}\)f_{\pm}^{(n)}(z) = m_n f_{\mp}^{(n)}(z),
\end{equation}
where ${\cal M}$ is a $2\times 2$ mixing matrix:
\begin{equation}
{\cal M}=
e^{-A}\left( 
\begin{array}{cc} 
M_L  & m(z) \\
m(z)& M_R 
\end{array} \right). 
\label{singlegenmatrix} 
\end{equation}
The equations for $f_+^{(n)}$ and $f_-^{(n)}$ can be decoupled by deriving a second-order equation
from (\ref{singlegenferm}). The fields $f_{\pm}^{(n)}$ obey a Schr\"{o}dinger-like equation:
\begin{equation}\label{secondorder}
\(-\partial_z^2+{\cal V}_{\pm}\)f_{\pm}^{(n)} = m_n^2 f_{\pm}^{(n)},
\end{equation}
where the ``potentials'' are given by:
\begin{equation}\label{potentials}
{\cal V}_\pm(z) = {\cal M}^2\mp{\cal M}'.
\end{equation}
The difficulty in solving \eqref{secondorder} is due to the fact that the mixing matrix generally cannot be diagonalized through global transformations of the functions, $f_{L,R\pm}^{(n)}$. However, there are special cases for which the second-order equations can be decoupled further. They occur whenever:
\begin{eqnarray}\label{cases}
M_L=M_R, &\qquad& \text{``degenerate''}\nonumber\\
M_L+M_R \pm \partial_z e^{A(z)} = 0. &\qquad& \text{``split''}
\end{eqnarray}
The ``degenerate'' case is separable in any background. The ``split'' cases are separable regardless of the Higgs VEV {\it in AdS}, where the split-case condition simply becomes $M_L+M_R \pm k=0$.

For generic forms of the Higgs VEV, it is most useful to work with transformed fields,
\begin{equation}\label{transf}
g_{\pm}^{(n)} = \left( 
\begin{array}{c} 
 g^{(n)}_{L\pm} \\
 g^{(n)}_{R\pm}
\end{array} \right)
= U f_{\pm}^{(n)}
\equiv \frac{1}{\sqrt{2}}
\left(\begin{array}{cc} 
 1 & 1 \\
 1 & -1
\end{array} \right)
\left(\begin{array}{c} 
 f^{(n)}_{L\pm} \\
 f^{(n)}_{R\pm}
\end{array} \right).
\end{equation}
In this basis the equations of motion are given by,
\begin{equation}\label{singlegenfermb}
\(\pm\partial_z+\widetilde{\cal M}\)g_{\pm}^{(n)}(z) = m_n g_{\mp}^{(n)}(z),
\end{equation}
where
\begin{equation}
\widetilde{{\cal M}}=U^{\dag}{\cal M}U=
\frac{e^{-A}}{2}\left( 
\begin{array}{cc} 
2m(z) + M_L + M_R  & M_L - M_R \\
M_L - M_R & -2m(z) + M_L + M_R
\end{array} \right),
\label{singlegenmatrixb} 
\end{equation}
while the boundary conditions \eqref{bcdir} become:
\begin{equation}
\label{bcprime0}
g^{(n)}_{L\pm} \big\vert_{z_0}= \pm g^{(n)}_{R\pm} \big\vert_{z_0}~.
\end{equation}
We may also define transformed potentials, $\widetilde{\cal V}_{\pm}$, in direct analogy with \eqref{potentials}. For the degenerate case, both of the potentials $\widetilde{\cal V}_+$ and $\widetilde{\cal V}_-$ are simultaneously diagonal in this basis. In the split cases, only one of the potentials $\widetilde{\cal V}_{\pm}$ will be diagonal. After solving the corresponding pair of decoupled second-order equations, the first-order equations \eqref{singlegenfermb} can then be used to generate the remaining solutions.

Below, we consider the degenerate case and one of the two split cases, $M_L+M_R+k=0$, assuming the following form for the Higgs VEV:
\begin{equation}
h(z)=\eta k^{3/2} \mu^2 z^2,   
\label{higgsvev}
\end{equation}
giving $m(z)=b k (\mu z)^2$ where $b=\lambda_5\eta/\sqrt{2}$, as in \cite{Batell:2008me}. We also parameterize the bulk masses in units of the AdS curvature, $M_{L,R}=c_{L,R} k$, where $c_{L,R}$ are dimensionless coefficients.

\subsubsection{Degenerate Bulk Masses}

The solution to the degenerate bulk mass case $c_L=c_R=c$ was presented in detail in \cite{Batell:2008me}. 
The lowest-lying mode was found to be:
\begin{equation}
m_0^2\simeq 
\begin{cases}
\frac{2 b \mu^2}{\Gamma(-1/2+|c|)}(b \mu^2 z_0^2)^{-1/2+|c|}~\qquad~{\rm for} \quad |c|>\frac{1}{2}~,\\
\frac{4 b \mu^2}{\pi \sec c\pi -\psi(1/2-c)-\psi(1/2+c)} \qquad\quad {\rm for} \quad |c| <\frac{1}{2}~,
\end{cases}
\end{equation}
where $\psi$ is the digamma function. Note that the lowest-lying mode is very light for  $\mu z_0 \ll1$ and 
$|c|>1/2$ (becoming exponentially small with increasing $|c|$), while for $|c| <1/2$ the fermion mass is of order 
$b\mu^2$.

\subsubsection{Split Localizations}

Here we consider one of the ``split'' cases, $c_L+c_R+1=0$. The other case, $c_L+c_R-1=0$, is very similar. With this choice, the transformation \eqref{transf} will diagonalize the potential $\widetilde{V}_+$ for any Higgs VEV in AdS. However, for our choice, $h(z)\sim z^2$, the untransformed potential $V_-$ happens to be diagonal,
\begin{equation}
\cal V_- = \(\begin{array}{cc}
\frac{c(c-1)}{z^2}+b^2 \mu^4 z^2& 0 \\
0 & \frac{(c+1)(c+2)}{z^2}+b^2 \mu^4 z^2
\end{array}\),
\end{equation}
so we will work in this basis.

A consistent solution requires that either $f_{L-}^{(n)}=0$ or $f_{R-}^{(n)}=0$. This is a peculiarity of the particular choice of the Higgs VEV and will not be true for other forms. The result is that the full tower of orthogonal solutions is most easily described in terms of two ``distinct'' KK towers of solutions. The first solution is:
\begin{eqnarray}
f_{L-}^{(n)}(z)&=&N_{L-}^{(n)}e^{-b\mu^2z^2/2}z^c ~U\(\frac{1}{4}+\frac{c}{2}-\frac{m_n^2}{4b\mu^2},\frac{1}{2}+c,b\mu^2z^2\),\label{flma}\\
f_{R+}^{(n)}(z)&=&\frac{b\mu^2 z}{m_n}f_{L-}^{(n)},\label{frpa}\\
f_{L+}^{(n)}(z)&=&\frac{1}{m_n} \(\frac{c}{z}f_{L-}^{(n)} - f_{L-}^{(n)'}\),\label{flpa}\\
f_{R-}^{(n)}(z)&=&0,\label{frma}
\end{eqnarray}
where $N_{L-}^{(n)}$ is a normalization constant and $U(a,b,y)$ is the Tricomi confluent hypergeometric
function. For this tower, the boundary conditions $f_{R+}^{(n)}\big|_{z_0}=0$ and $f_{L-}^{(n)}\big|_{z_0}=0$ are equivalent. There is only a single orthonormality condition,
\begin{equation}
\int_{z_0}^{\infty}dz\,f_{L-}^{(n)}f_{L-}^{(m)} =\delta^{nm},
\end{equation}
which in fact implies the correct orthonormality condition for the remaining fields,
\begin{equation}
\int_{z_0}^{\infty} dz\,\[f_{L+}^{(n)}f_{L+}^{(m)}+f_{R+}^{(n)}f_{R+}^{(m)}\] =\delta^{nm}.
\end{equation}
The other KK tower is given by:
\begin{eqnarray}
f_{R-}^{(n)}(z)&=&N_{R-}^{(n)}e^{-b\mu^2 z^2/2}z^{2+c}~U\(\frac{5}{4}+\frac{c}{2}-\frac{m_n^2}{4b\mu^2},\frac{5}{2}+c,b\mu^2 z^2\),\label{frmb}\\
f_{L+}^{(n)}(z)&=&\frac{b\mu^2 z}{m_n}f_{R-}^{(n)},\label{flpb}\\
f_{R+}^{(n)}(z)&=&-\frac{1}{m_n} \(\frac{1+c}{z}f_{R-}^{(n)} +f_{R-}^{(n)'}\),\label{frpb}\\
f_{L-}^{(n)}(z)&=&0,\label{flmb}
\end{eqnarray}
where $N_{R-}^{(n)}$ is a normalization constant. The spectrum for this tower is found by imposing the boundary condition $f_{R+}^{(n)}\big|_{z_0}=0$, while the normalization condition is,
\begin{equation}
\int_{z_0}^{\infty}dz\,f_{R-}^{(n)}f_{R-}^{(m)}=\delta^{nm}.
\end{equation}
The lowest lying mode of the second tower is very light and approximating the mass requires some care. For 
$m_0^2\ll b\mu^2$, we can expand the functions using techniques of so-called boundary perturbation theory of quantum mechanics \cite{Hull:1956,Gorecki:1987}. The function $f_{R-}^{(n)}$ obeys the Schr\"{o}dinger-like equation,
\begin{equation}\label{schrodinger}
-\partial_z^2 f_{R-}^{(n)}+\[\frac{(c+1)(c+2)}{z^2}+b^2\mu^4 z^2\]f_{R-}^{(n)} = m_n^2 f_{R-}^{(n)}.
\end{equation}
For small $m_0^2$, we can write $f_{R-}^{(0)}$ as a product of the zero-mode solution and a correction:
\begin{equation}
f_{R-}^{(0)}=\zeta(z)F(z),
\end{equation}
where $\zeta(z)$ satisfies the zero-mode equation
\begin{equation}\label{zmode}
-\partial_z^2 \zeta+\[\frac{(c+1)(c+2)}{z^2}+b^2\mu^4 z^2\]\zeta = 0.
\end{equation}
The solution may be written as,
\begin{equation}
\zeta(z)=N_{R-}^{(0)}{z}^{1/2}K_{\nu}(b\mu^2z^2/2),
\end{equation}
where $N_{R-}^{(0)}$ is a constant, $K_\nu$ is the modified Bessel function and $\nu=3/4+c/2$. The function $F(z)$ obeys the second-order equation,
\begin{equation}
\[\zeta^{-2}\partial_{z}(\zeta^2\partial_{z})+m_0^2\]F(z)=0,
\end{equation}
and may be expanded in powers of $m_0^2$ as,
\begin{equation}
f_{R-}^{(0)}\simeq\zeta(z)\[1+m_0^2\int_{z_0}^{z}dz'\,\zeta^{-2}\int_{z'}^{\infty}dz''\,\zeta^2+\mathcal{O}(m_0^4)\].
\end{equation}
Such an expansion has also been used in Ref.\cite{Falkowski:2008fz} to approximate wavefunctions in soft-wall models. In contrast, here we are using it to solve the boundary value problem. The UV boundary condition,
\begin{equation}\label{bcsplit}
f_{R+}^{(0)}\Big|_{z_0}=\(f_{R-}^{(0)'}+\frac{1+c}{z}f_{R-}^{(0)}\)\Big|_{z_0}=0,
\end{equation}
results in the following approximate expression for $m_0^2$:
\begin{equation}
m_0^2\simeq\frac{K_{1-\nu}(b\mu^2 z_0^2/2)}{K_{\nu}(b\mu^2 z_0^2/2)\mathcal{I}(z_0)}z_0\label{massform},
\end{equation}
where
\begin{equation}
\mathcal{I}(z)\equiv\int_{z}^{\infty}dz'\,\zeta^2(z').
\end{equation}
The expression (\ref{massform}) can now be expanded for small $z_0$. For $c > -1/2$ we find,
\begin{equation}
m_0^2 \simeq \frac{2(c/2+3/4)}{\Gamma\(c/2+3/4\)}\(\frac{b^2 \mu^4 z_0^2}{4}\)\[\(\frac{b \mu^2 z_0^2}{4}\)^{c-1/2}\Gamma\(-c/2+1/4\)+\Gamma\(c/2-1/4\)\].
\label{simplemass}
\end{equation}
In the limit $c\gg 1/2$, this expression simplifies further to
\begin{equation}\label{smallsplitmass}
m_0^2\simeq\(1+\frac{1}{c-1/2}\){b^2\mu^4} z_0^2.
\end{equation}
This expression reveals a lower bound on the fermion mass in this region of the bulk mass parameter space, $m_{\rm min} = (\mu z_0)b\mu$.

\begin{figure}[t]
\centerline{\includegraphics{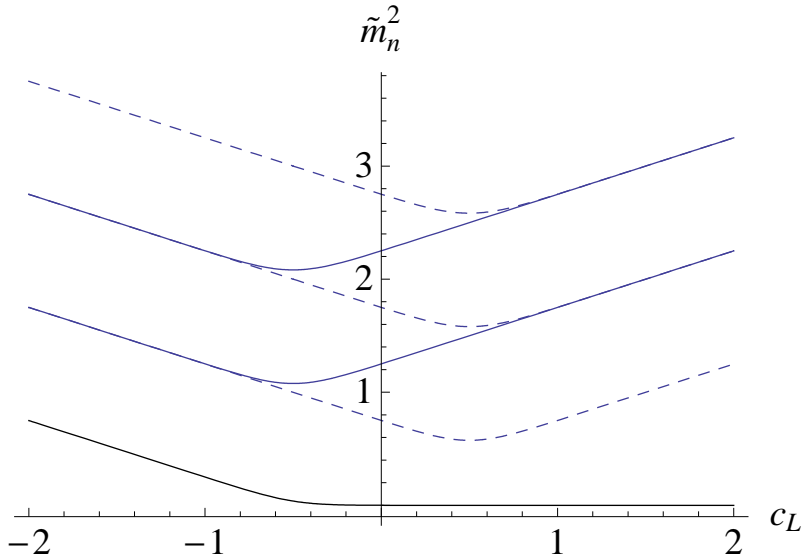}}
\caption{The first several masses in the split case plotted as a function of $c_L$. The separate KK towers (dashed 
and solid lines) coincide for large negative and positive values of $c_L$. Here, $\tilde{m}_n^2=m_n^2/b\mu^2$ and 
$\mu z_0\ll 1$.\label{towerplot}}
\end{figure}

For $c\ll -1/2$, the above expansions are poor approximations because the mass becomes $\mathcal{O}(b\mu)$. To deal with this regime, we can apply mathematical techniques from ``supersymmetric quantum mechanics'' to determine the mass \cite{Cooper:2001zd}. Consider a quantum mechanical system for which the Hamiltonian may be factorized as:
\begin{equation}
\left[-\partial_z + W(z)\right]\left[\partial_z+W(z)\right]\psi = m_n^2 \psi.
\label{fosusyqm}
\end{equation}
The ``superpotential,''
\begin{equation}
W(z)=\frac{1+c}{z}+b\mu^2 z,
\end{equation}
gives rise to the ``ordinary'' potential for the function $\psi$:
\begin{equation}
V(z)=W^2-W'=\frac{(c+1)(c+2)}{z^2}+b^2\mu^4z^2+(2c+1)b\mu^2.\label{shifted}
\end{equation}
It is clear from (\ref{fosusyqm}) that there exists a zero mode solution, $\psi\sim e^{-\int W}$, with boundary conditions that are given trivially by the equations of motion. In the limit $\mu z_0 \to 0$, however, this is equivalent to the boundary condition \eqref{bcsplit}. Since the potential in \eqref{schrodinger} is equivalent to \eqref{shifted} up to a constant shift of the reference potential, we can conclude then that the solution $\psi\sim e^{-\int W}$ is in fact a good approximation for $f_{R-}^{(0)}$ and that,
\begin{equation}\label{largesplitmass}
m_0^2\simeq(-1-2c)b\mu^2.
\end{equation}
This is clearly only valid when $c < -1/2$. We have checked that the expressions (\ref{smallsplitmass}) and (\ref{largesplitmass}) match the exact results well in this region. The full spectrum is plotted in Fig. \ref{towerplot}. The distinctive KK tower structure of the split case suggests the possibility of novel KK physics unlike that found in hard-wall models and may be interesting to study in other soft-wall bulk Higgs models as well.

\subsection{Comparison with Perturbative Expansions}

Recently, the possibility of modeling fermions by introducing non-constant bulk Dirac mass terms has been considered in Ref.\cite{Aybat:2009mk}. For a single generation setup with quadratic bulk mass terms, the equations of motion are the same as \eqref{singlegenferm}, but now the mixing matrix \eqref{matrix} can be written effectively as,
\begin{equation}
{\cal M}=
e^{-A}\left( 
\begin{array}{cc} 
c_L^0 k + c_L^1 k \mu^2 z^2 & b k \mu^2 z^2 \\
b k \mu^2 z^2& c_R^0 k + c_R^1 k \mu^2 z^2 
\end{array} \right),
\label{matrixb}
\end{equation}
where $c_{L,R}^0, c_{L,R}^1$ are constant coefficients. The effect of this non-constant bulk mass is that normalizable zero-modes persist (depending on the choice of the signs of $c_{L,R}^1$) even in the limit $b\to 0$. For small values of $b$, the spectrum may be found by treating the bulk Yukawa interaction as a perturbation on the $b=0$ solutions.

Such an approach can be related to ours in some cases. For example, in the case of degenerate constant mass pieces, $c_L^0=c_R^0$, global unitary transformations may still be used to diagonalize the mass matrix when the bulk masses have the same functional form as the Higgs VEV. Thus, the introduction of non-constant bulk masses can be viewed as effectively changing the boundary conditions on the fields in such cases.

It is interesting to note that the case considered in \cite{Aybat:2009mk} is similar to our ``split'' case. In particular, they examine $c_L^1=-c_R^1$ and $c^0_L= - c^0_R$ in detail. In the slightly different split configuration for the constant pieces of the bulk mass, $c_L^0 = \pm 1-c_R^0$, analytical solutions can be obtained in a similar fashion to our earlier analysis. For $c_L^0=-1-c_R^0$, we find the lowest lying KK tower to be:
\begin{eqnarray}
f_{R-}^{(n)}(z)&=&N_{R-}^{(n)}e^{-{\tilde{b}\mu^2 z^2/2}}z^{2+c_L^0}~U\(\frac{5}{4}+\frac{c_L^0}{2}-\frac{\tilde{m}_n^2}{4\tilde{b}\mu^2},\frac{5}{2}+c_L^0,\tilde{b}\mu^2 z^2\),\label{frmbp}\\
f_{L+}^{(n)}(z)&=&\frac{b\mu^2 z}{m_n}f_{R-}^{(n)},\label{flpbp}\\
f_{R+}^{(n)}(z)&=&-\frac{1}{m_n} \(f_{R-}^{(n)'}+ \frac{1+c_L^0}{z}f_{R-}^{(n)} +c_L^1 \mu^2 z f_{R-}^{(n)}\),
\label{frpbp}\\
f_{L-}^{(n)}(z)&=&0,\label{flmbp}
\end{eqnarray}
where we have defined effective parameters, $\tilde{b}^2 = \(c_L^1\)^2 + b^2$, and 
$\tilde{m}_n^2 = m_n^2 - c_L^1\(2c_L^0+1\)$ to make the comparison with (\ref{frmb})-(\ref{flmb}) clear. 
Note that there remains a lower bound on the mass for $c_L^0\gg 1/2$. We have checked that this solution describes the large $c_L^0$ behavior for the case considered in Ref.\cite{Aybat:2009mk}, $c_L^0 = -c_R^0$. We expect that all of the basic features of the non-constant bulk mass model should be contained within our exact solutions.

Generically the split and degenerate cases allow one to find the spectrum exactly by solving a set of decoupled second-order equations. Even when the equations cannot be solved exactly, approximate methods such as those we have described above may be employed. Additionally, as our numerical results will verify, one can expect the behavior in these special cases to provide a complete qualitative picture of the full parameter space dependence.

\subsection{Couplings to Gauge Bosons}\label{gauge}
Of significant interest in models involving extra dimensions is the coupling of fermions to the KK gauge bosons. When the fermions are localized at different points along the extra dimension, they can obtain non-universal couplings to the excited gauge bosons. Such non-universality will generically lead to large contributions to flavor physics observables, providing very stringent lower bounds on the allowed KK scale \cite{Delgado:1999sv}. 

In hard-wall models, the couplings can become universal for certain regions of the parameter space, resulting in a GIM-like suppression of flavor changing neutral currents \cite{Gherghetta:2000qt}, thereby greatly lowering the bound on the allowed KK scale. We therefore would like to see if a similar effect is present in the soft-wall case. Moreover, we would like to develop our formalism in such a way that multiple fermion generations can be incorporated.

The couplings of the zero mode fermions to the KK gauge bosons are found to be:
\begin{equation}\label{kkgauge}
g^{\alpha\beta(n)}_{\pm} = g_5 \int_{z_0}^{\infty}dz\, f_A^{(n)} \[ (f_{L\pm}^{i\alpha(0)})^{\dag}f_{L\pm}^{i\beta(0)}+ 
(f_{R\pm}^{i\alpha(0)})^{\dag}f_{R\pm}^{i\beta(0)}\],
\end{equation}
where $f_A^{(n)}$ is the gauge boson profile along the extra dimension. The gauge boson profiles arising from a quadratic dilaton \eqref{dilaton} were derived in \cite{Batell:2008me}. Here we simply use the results. The zero-mode couplings $g^{\alpha\beta(0)}_\pm\equiv g\delta^{\alpha \beta}$ remain universal due to the orthonormality condition \eqref{normf} and the flat zero mode gauge boson profile. This is because the dilaton factor explicitly cancels and plays no role.

\begin{figure}
\centerline{\includegraphics{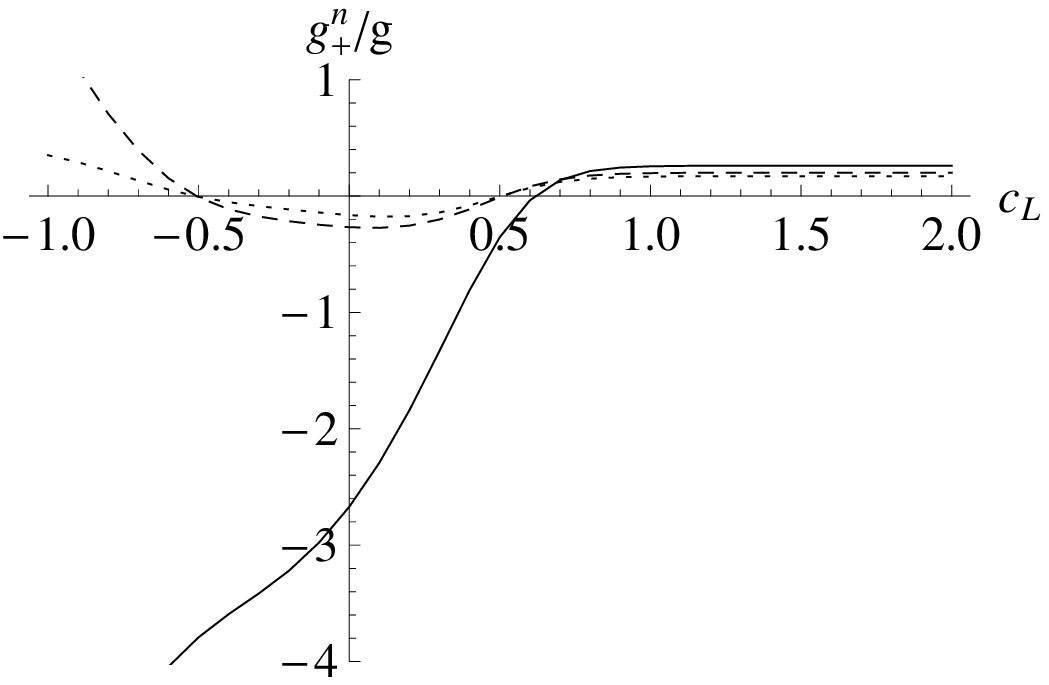}}
\caption{The ratio $g^n_+/g$ for $n=1$ (solid), $n=2$ (dashed), and $n=3$ (dotted) KK gauge modes coupled to 
the zero-mode fermion in the split case, as calculated using \eqref{kkgauge}.}
\label{gaugeplota}
\end{figure}

\begin{figure}
\centerline{\includegraphics{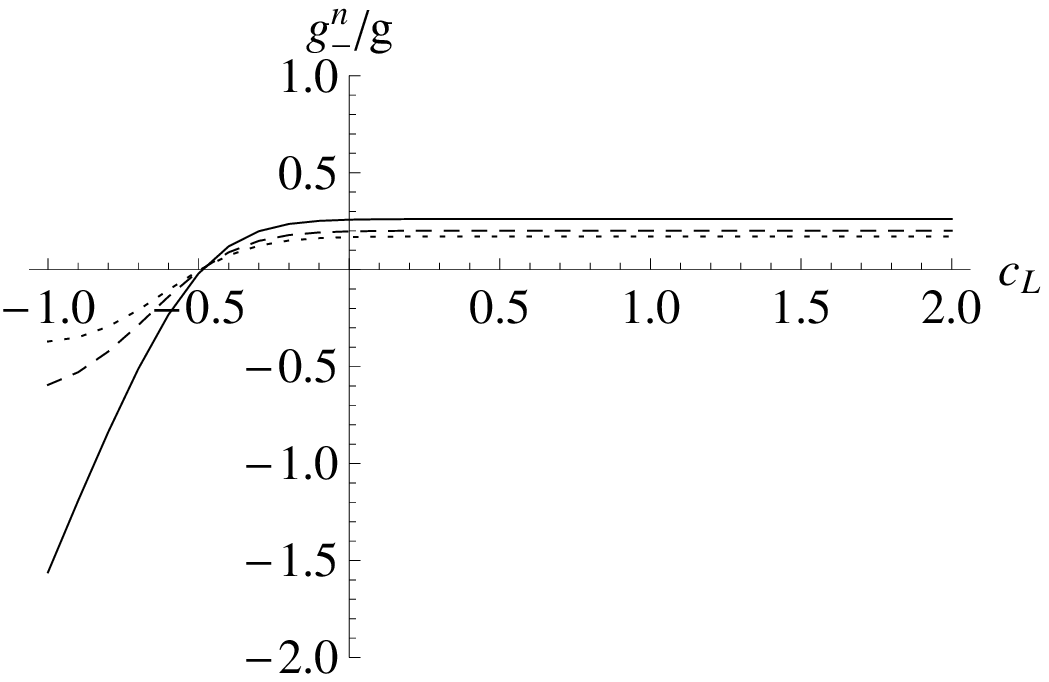}}
\caption{The ratio $g^n_-/g$ for $n=1$ (solid), $n=2$ (dashed), and $n=3$ (dotted) KK gauge modes coupled to 
the zero-mode fermion in the split case, as calculated using \eqref{kkgauge}.}
\label{gaugeplotb}
\end{figure}

The degenerate single-generation case was considered in \cite{Batell:2008me}, where it was found that only one of the couplings, $g_{+}$ or $g_{-}$, can become universal due to the opposite localizations of the fermion modes. In Ref \cite{Aybat:2009mk}, it was seen that opposite constant and non-constant bulk masses led to universal couplings for both $g_{+}$ and $g_{-}$. This happens as well for the split case solutions. We have plotted the couplings for this case in Figs. \ref{gaugeplota} and \ref{gaugeplotb}. We find that the couplings $g_+$ and $g_-$ become universal simultaneously whenever $c\gg 1/2$.

Note that the bounds from flavor physics are generically expected to be more stringent in soft-wall models than in models with a hard wall. This follows from the generically closer spacing of the KK modes in soft-wall models as compared to hard-wall models. For example, we can consider the contribution to $\Delta m_K$ arising from non-universal couplings. The effective 4D Lagrangian contains operators that are suppressed by the squared masses 
of the KK gauge bosons mediating the strangeness-changing transitions ($\Delta S=2$):
\begin{equation}\label{flavorops}
\mathcal{L}_{\Delta S=2} \supseteq \sum_{n=1}^{\infty} \frac{1}{M_n^2}\[\bar{d}_{L}^{\alpha}\tilde{g}_{+}^{\alpha\beta(n)}\gamma^{\mu} d_{L}^{\beta}+ \bar{d}_{R}^{\alpha}\tilde{g}_{-}^{\alpha\beta(n)}\gamma^{\mu}d_{R}^{\beta}+{\rm~h.c.}\]^2,
\end{equation}
where the sum is over the gauge boson KK modes with KK masses $M_n$, and $\tilde{g}^{\alpha\beta(n)}_{\pm}=V_{L,R}^d\,g^{\alpha\beta(n)}_{\pm}V_{L,R}^{d\dagger}$ with $V_{L,R}^d$ generic unitary 
matrices~\cite{Delgado:1999sv}. Thus, in the presence of non-degenerate couplings to the bulk KK gauge bosons, bounds from flavor experiments may be interpreted as a lower bound on the KK scale.

The key point is that the total amount of suppression in \eqref{flavorops} depends upon the spacing of the KK tower. In a hard-wall model, for example, $m_n^2\sim n^2 M_{KK}^2$, where $M_{KK}$ is the KK mass scale. This compares with the soft-wall scenario where it would seem to imply a problem, because the squared mass trajectories grow generically as $m_n^2 \sim n M_{KK}^2 $ (indeed, this spacing was the original motivation for studying the soft-wall \cite{Karch:2006pv}). While the sum of $1/n$ diverges as $n\to\infty$, we should of course truncate the sum at some high energy cutoff. Nevertheless, the naive implication is that the constraints on soft-wall models should be considerably tighter.

However, this argument ignores the fact that the gauge bosons become increasingly IR localized with increasing mode number $n$. Thus, any off-diagonal terms in the gauge coupling matrices are further suppressed for large $n$.
By performing a numerical fit using the first several dozen gauge boson modes and our split case solutions, we find that the couplings fall off as $n^{-0.4}$ to a very good approximation in the region where the couplings are independent of localization. This implies that the terms in the sum \eqref{flavorops} grow as $n^{-1.4}$. All other things being equal, this implies that the constraints from flavor physics are roughly a factor of two more stringent in this model than in hard-wall models.

While this presents no great problem for the model with a {\it quadratic} dilaton, for a generic power law behavior in the dilaton, $\Phi\sim z^{\alpha}$, the spectrum of gauge bosons grows as $m_n^2\sim n^{2-2/\alpha}M_{KK}^2$ \cite{Batell:2008me}. This means that for less steep potentials, even tiny amounts of non-degeneracy among the bulk couplings has potentially severe implications for flavor physics.

\begin{figure}[t]
\centerline{\includegraphics[width=.6\textwidth]{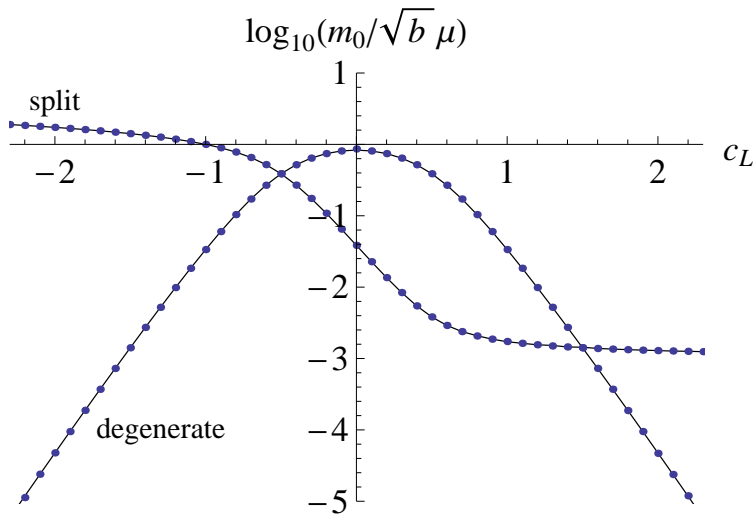}}
\caption{Lowest lying masses for the ``degenerate'' $(c=c_L=c_R)$ and ``split" $(c=c_L=-1-c_R)$ cases, where the solid lines are determined in Ref.\cite{Batell:2008me} (degenerate) and from \eqref{massform} (split). The dots represent values obtained using the numerical method of Section \ref{routine}.}
\label{comparisonplot}
\end{figure}

\section{Numerical Solution}\label{numerical}

\subsection{Routine}\label{routine}

The analytical solutions that we have presented are of limited use, and instead we would like to solve the full fermion mass problem including flavor. Our goal is to solve the eigenvalue problem \eqref{genferm} with mixed boundary conditions. The ``initial conditions'' \eqref{bcdir} specify half of the boundary values at the UV brane or, equivalently, half of the integration constants for the system. The remaining constants of integration are fixed by the normalization conditions \eqref{normf}, which can only be satisfied if the eigenvalue, $m_n^{\alpha}$, has been chosen correctly.

We convert the problem to an initial value one by extending the shooting method to linear boundary value problems of arbitrary order \cite{Keller:1976}. The solutions to \eqref{genferm} may be written as:
\begin{equation}\label{prop}
f^{i \alpha (n)}(z)=U(m_n^{\alpha}; z,z_0)^{i j}f^{j \alpha (n)}(z_0),
\end{equation}
where the propagator $U(m_n^{\alpha}; z,z_0)$ is a linear operator and the $f^{i \alpha (n)}(z)$ are $4N_F\times 4N_F$ matrix-valued functions for $N_F$ fermion generations.

The matrix elements of $U$ may be found by integrating a set of $4N_F$ linearly independent basis vectors that span the space of initial values, $f^{i \alpha (n)}(z_0)$, and inverting \eqref{prop}. The initial values that lead to normalizable solutions correspond to eigenvectors of $U(m_n^{\alpha}; z,z_0)$ with vanishing eigenvalues in the limit $z\to\infty$. There are generally $2N_F$ such eigenvectors. Numerically, we can estimate the values of these vectors by considering the eigenvectors of $U(m_n^{\alpha}; z_1,z_0)$, where our cutoff satisfies $z_1\gg \mu^{-1}$. In practice, results are much more reliable if one starts the forward integration from some intermediate range $z^{*}\sim\mu^{-1}$ and then integrates the normalizable modes back to $z_0$. Variations on this theme can be explored.

We scan over $m_n^{\alpha}$, at each point integrating the system from a set of initial values so as to reconstruct the normalizable solutions. If there exists a linear combination of the solutions that matches the boundary conditions \eqref{bcdir}, then $m_n^{\alpha}$ is a solution to the system. To determine when this occurs, we define a merit function as the absolute value of the determinant of a matrix and search for a minimum. The matrix we use has columns formed by projecting out of the normalizable initial value vectors those components that are not restricted by the boundary conditions. These projected column vectors must be linearly dependent in order to satisfy the boundary conditions of the problem.

When the hierarchy between $\mu$ and $z_0^{-1}$ is very large, increasingly high precision is necessary to achieve reliable results. Iterative methods may be better suited to the problem in such cases. Our primary goal is to highlight the differences between fermions in soft-wall and hard-wall scenarios, and the speed and simplicity of this technique are its chief advantages. For this reason, we have limited our attention to a modest hierarchy.

\subsection{Results}\label{results}
\subsubsection{Single Generation}

\begin{figure}
\centerline{\includegraphics[width=.55\textwidth]{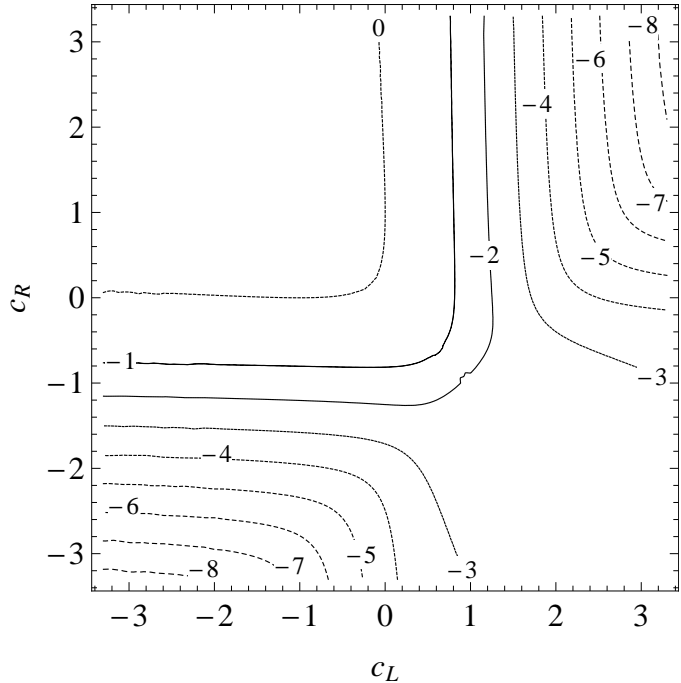}}
\caption{Contours of $\log_{10}(m_0/\sqrt{b}\mu)$ for the lowest lying masses in our soft-wall setup
with $\sqrt{b}\mu z_0=10^{-3}$.}
\label{contours}
\end{figure}

We first present results for a single generation of fermions, as this case illustrates the essential features of the 
fermion mass behavior in the soft-wall, and allows us to compare our numerical results with the analytical cases 
in the appropriate limits as well as to a typical hard-wall setup. Assume the following values of the parameters:
\begin{equation}
\mu = 1 ~{\rm TeV}; \qquad \mu z_0 = 10^{-3}; \qquad b = 1.
\end{equation}
In Figure \ref{comparisonplot}, we compare the numerical results to the analytical results from Section \ref{singlegensection} where it can be seen that the two methods agree very well. Next in Figure \ref{contours}
we plot the fermion mass contours to show the full dependence on the parameters $c_L$ and $c_R$ . The shape of the plot is easily understood from the analytical results. The numerical solution smoothly interpolates between the solutions along the lines $c_L=c_R$ and $c_L=\pm 1-c_R$. Because a similar analysis can be repeated for other Higgs VEVs, this provides a natural way to begin studying the qualitative aspects of other models in AdS as well.

\begin{figure}
\centerline{\includegraphics[width=.55\textwidth]{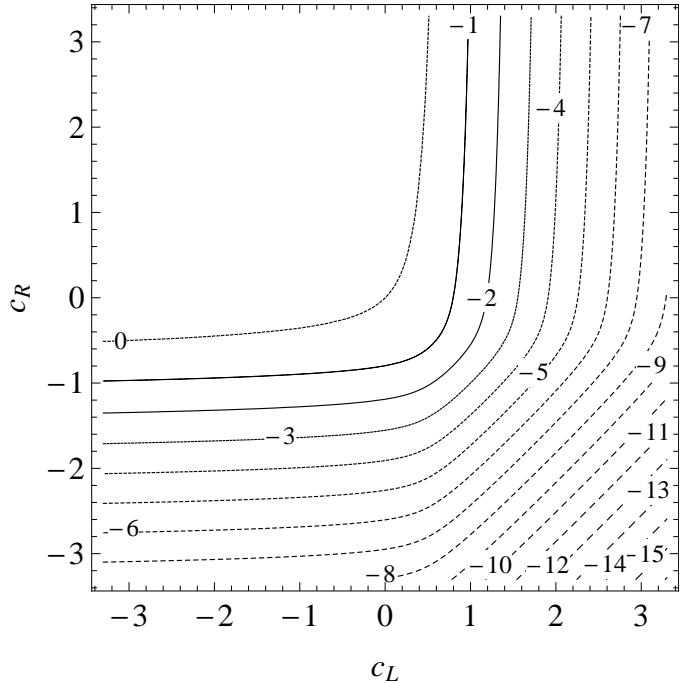}}
\caption{Contours of $\log_{10}(m_0/\sqrt{b}\mu)$ for the lowest lying masses in a typical hard-wall setup
with $\sqrt{b}\mu z_0=10^{-3}$.}
\label{hardwall}
\end{figure} 

We can compare the soft-wall behavior with a typical hard-wall setup. In Figure \ref{hardwall} we provide the corresponding contour plot for a hard-wall model in which the SM fermion masses are simply proportional to the values of the wavefunctions at $z=1/\mu$. The most striking difference between the plots occurs in the lower right-hand corner. This is the region where $c_L>1/2$ and $c_R<-1/2$.

The hard-wall case is characterized by a steep dependence on the bulk mass in this region, where the wavefunctions are proportional to $f_{L+}^{(0)}\sim z^{-c_L}$ and $f_{R-}^{(0)}\sim z^{c_R}$. For $z_0\ll \mu^{-1}$, the normalization constants become vanishingly small:
\begin{equation}
N_{L,R}^{(0)}\sim z_0^{-1/2\pm c_{L,R}}.
\end{equation}
Thus, the values of the functions in the IR at $z=\mu^{-1}$ are additionally suppressed. This is the well-known mechanism for generating SM mass hierarchies in Randall-Sundrum scenarios with bulk fields~\cite{Grossman:1999ra,Gherghetta:2000qt}. For the soft-wall case, however, we can see the lower bound on the mass in this region,
\begin{equation}
m_0\sim (\mu z_0)\mu,
\end{equation}
as indicated by the approximate expression \eqref{smallsplitmass}. This can be understood be noting
that the normalization (\ref{normf}) involves the sum of two types of fermion contributions which are 
generically not simultaneously suppressed.

\subsubsection{Three Generations}

Next we aim to provide concrete numerical examples involving three generations of fermions that fully take into account the 5D flavor mixing to show that the attractive features of the soft-wall are maintained.
For multiple generations, there are three matrices that parameterize the fermions: two bulk mass matrices $M_L$ and $M_R$, and the bulk Yukawa matrix, $\lambda_5$. We take the action \eqref{ferma} to be written in an arbitrary basis, for example, the CKM basis. Absent some symmetry, there is no reason to expect any structure relating the entries of the various bulk parameter matrices. We generically expect that the entries of each matrix are all of order unity (in units of the AdS curvature scale, $k$), and that the various matrices are misaligned. There is of course some basis in which both $M_L$ and $M_R$ are diagonal. Thus, by ``misaligned,'' we mean that this basis is distinct from the one in which the Yukawa matrix is diagonal. Indeed, the typical approach is to work in this basis, treating the Yukawa interactions as perturbations. Such an approach has been used in both hard-wall \cite{Grossman:1999ra, Gherghetta:2000qt, Huber:2000ie, Grossman:2004rm,Casagrande:2008hr} and even very recently in soft-wall setups \cite{Aybat:2009mk}.

In Ref.\cite{Aybat:2009mk}, it was found that one needed to include the first several ($\sim 10$) KK modes in order to achieve reliable results in such a perturbative expansion when including only a single generation. At such a point, the analysis is essentially a numerical exercise. In our view, it is advantageous to include the entire KK tower in the numerical formulation wherever possible. In other words it may make the most sense to simply solve the equations of motion \eqref{genferm}, which guarantee the orthogonality of the eigenfunctions due to the hermiticity of the mixing matrix.

We expect that all other interactions may be treated reliably as perturbations. This is because the Higgs grows unbounded in the IR where it is the dominant contribution to the fermion equations of motion. Other observables may thus be calculated using the usual wavefunction overlap approximation. As an application, we will calculate the couplings to excited gauge bosons for examples involving three generations.

We do not attempt to set precise bounds on soft-wall models here, as doing so goes significantly beyond the scope of this work. Electroweak and flavor constraints have been discussed in the context of soft-wall models in Refs. \cite{Batell:2008me,Aybat:2009mk}. Detailed analyses in various hard-wall scenarios can be found in \cite{Delgado:1999sv,Grossman:1999ra,Gherghetta:2000qt,Huber:2000ie,Csaki:2008zd,Casagrande:2008hr} and references therein.

However, we will require that the eigenvalues of the bulk mass matrices satisfy $m_L^i\gtrsim k/2$ and $m_R^i \lesssim -m_L^i$ in order to get nearly degenerate gauge couplings. Because of the lower bound on the fermion masses at $m_0\sim (\mu z_0)\mu$ in this region, it is clear that the hierarchy considered above, $\mu z_0 = 10^{-3}$ will be inadequate for generating MeV scale masses when $\mu=1~{\rm TeV}$, and will only be possible for 
$\mu z_0 \lesssim 10^{-6}$. Thus we again assume a quadratic Higgs VEV, $h(z)=\eta k^{3/2} \mu^2 z^2$, and 
the following for our input parameters:
\begin{equation}
\mu = 1~{\rm TeV};\qquad \mu z_0 = 10^{-6}.
\label{newvals}
\end{equation}
Dealing with much larger hierarchies presents significant numerical challenges. However, the qualitative results of such an analysis should not be substantially different from the results presented here.

\begin{figure}
\centerline{\includegraphics[width=.7\textwidth]{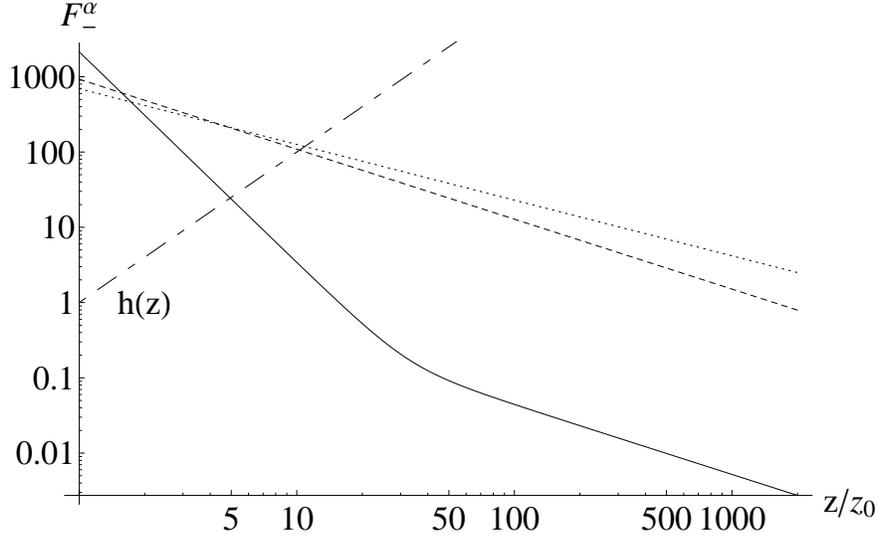}}
\caption{The down-type fermion bulk profiles $F_-^\alpha(z)$ (in units of $\sqrt{\mu}$) for the first generation (solid), second generation (dashed) and third generation (dotted) showing the overlap with the Higgs VEV $h(z)$ (in units of
$\mu^2/\sqrt{k}$ with $\eta=1$).}
\label{Mdplot}
\end{figure}

First, we present an example resembling down-type quarks (or charged leptons). For simplicity, we take the entries of $M_L$ to be nearly degenerate, but we allow for large non-degeneracy in the matrix $M_R$ as well as in the Yukawa matrix. Specifically, we consider,
\begin{eqnarray}
\frac{M_L}{k} =\(\begin{array}{ccc}
0.784 & -0.020 & 0.023 \\
-0.020 & 0.808816 & 0.0094 \\
0.024 & 0.0094 & 0.780
\end{array}\), &~& 
\frac{M_R}{k} =\(\begin{array}{ccc}
-2.179 & -0.459 & -0.774 \\
-0.459 & -1.073 & -0.354 \\
-0.774 & -0.354 & 0.1218
\end{array}\),\nonumber
\end{eqnarray}
\begin{equation}
\frac{\eta}{\sqrt{2}}\lambda_5 =\(\begin{array}{ccc}
0.422 & 0.175 & -0.678 \\
-0.007 & 0.928 & 0.348 \\
0.295 & -0.327 & 0.637
\end{array}\).
\label{5Ddownvar}
\end{equation}
We find a spectrum of masses resembling the down-type quarks (or charged leptons):
\begin{equation}
m_0^{\alpha} = 0.57~{\rm MeV}, ~96.08~{\rm MeV}, ~1.310~{\rm GeV}.
\end{equation}
The fermion mass hierarchy is clearly obtained, but due to the complexity of the numerical procedure 
we do not match the SM masses exactly, and postpone a more detailed analysis for future work.
The fermion bulk profiles, $F_-^\alpha(z)=\sqrt{ (f_{L-}^{i\alpha(0)})^{\dag}f_{L-}^{i\alpha(0)}+ (f_{R-}^{i\alpha(0)})^{\dag}f_{R-}^{i\alpha(0)}}$  are plotted in Figure~\ref{Mdplot}. The fermion profile overlap with the Higgs VEV, $h(z)$ leads to the fermion mass 
hierarchy. The corresponding bulk profiles, $F_+^\alpha(z)$ are not plotted because the profile differences 
between the flavors are not as pronounced. This is due to our choice of UV boundary conditions and 
bulk masses (\ref{5Ddownvar}).

From expression \eqref{kkgauge}, we can calculate the coupling of the zero mode fermions to the KK gauge bosons (i.e. gluons). The result is a matrix whose off-diagonal entries contribute to flavor violation. We obtain the following results for the first two KK gauge coupling matrices, normalized to the coupling to the massless gauge boson:
\begin{eqnarray}
\frac{g^{(1)}_+}{g} =
\(\begin{array}{ccc}
0.186 &  10^{-4} & 10^{-4} \\
 10^{-4} & 0.187 & 2\times10^{-4} \\
 10^{-4} & 2\times10^{-4} & 0.185
\end{array}\),&&
\frac{g^{(2)}_+}{g} = 
\(\begin{array}{ccc}
0.140 &  10^{-4} &  10^{-4} \\
 10^{-4} & 0.138 &  10^{-4} \\
 10^{-4} &  10^{-4} & 0.137
\end{array}\);\nonumber\\
\frac{g^{(1)}_-}{g} =
\(\begin{array}{ccc}
0.188 & \approx 0 & \approx 0 \\
\approx 0 & 0.188 &  10^{-4} \\
\approx 0 &  10^{-4} & 0.184
\end{array}\),&&
\frac{g^{(2)}_-}{g} = \(\begin{array}{ccc}
0.139 & \approx 0 & \approx 0 \\
\approx 0 & 0.139 &  10^{-4} \\
\approx 0 &  10^{-4} & 0.137
\end{array}\).
\end{eqnarray}
This behavior is maintained for higher modes as well. For this choice of parameters, the very nearly degenerate couplings imply that $\mu$ of order a few TeV will be consistent with flavor constraints \cite{Csaki:2008zd,Casagrande:2008hr, Delgado:2009xb}. Note that we have assumed no contributions to CP violation. Thus the soft-wall model can accommodate the fermion mass hierarchy with large bulk mixing and small flavor violation.

\begin{figure}
\centerline{\includegraphics[width=.7\textwidth]{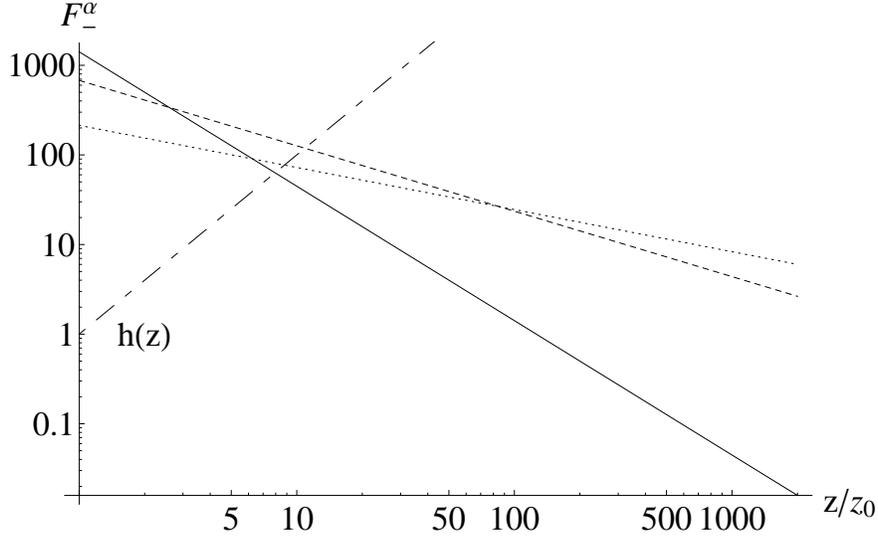}}
\caption{The up-type fermion bulk profiles $F_-^\alpha(z)$ (in units of $\sqrt{\mu}$) for
the first generation (solid), second generation (dashed) and third generation (dotted) showing the
overlap with the Higgs VEV $h(z)$ (in units of $\mu^2/\sqrt{k}$ with $\eta=1$).}
\label{Muplot}
\end{figure}

The up-type quarks are only moderately more sensitive to the presence of the top quark when large bulk mixing is allowed. For the choices (\ref{newvals}) we obtain 
\begin{eqnarray}
\frac{M_L}{k} =
\(\begin{array}{ccc}
0.749 & -0.005 & 0.017 \\
-0.005 & 0.785 & 0.066 \\
0.017 & 0.066 & 0.516
\end{array}\), &~& 
\frac{M_R}{k} =
\(\begin{array}{ccc}
-0.940 & -0.285 & -0.200 \\
-0.285 & -1.103 & -0.338 \\
-0.200 & -0.338 & -0.657
\end{array}\),\nonumber
\end{eqnarray}
\begin{equation}
\frac{\eta}{\sqrt{2}}\lambda_5 =
\(\begin{array}{ccc}
0.700 & -0.352 & -0.193\\
-0.079 & 0.826 & -0.065 \\
-0.098 & -0.321 & 1.430
\end{array}\),
\label{5Dupvar}
\end{equation}
which gives rise to the following mass spectrum:
\begin{equation}
m_0^{\alpha} = 2.10~{\rm MeV}, ~129.1~{\rm MeV}, ~151.5~{\rm GeV}.
\end{equation}
Again we see that the correct fermion mass hierarchy can be obtained.
The fermion bulk profiles, $F_-^\alpha(z)$  are plotted in 
Figure~\ref{Muplot}. The fermion profile overlap with the Higgs VEV, $h(z)$ leads to the fermion mass 
hierarchy. Similarly to the down-type fermions, the corresponding up-type 
bulk profiles $F_+^\alpha(z)$ are not plotted because the profile differences are negligible
due to the choice of UV boundary conditions and bulk masses (\ref{5Dupvar}).
The gauge couplings are nearly universal among the first two generations:
\begin{eqnarray}
\frac{g^{(1)}_+}{g} =\(\begin{array}{ccc}
0.186 & 2\times 10^{-3} &  2\times10^{-3} \\
2\times 10^{-3} & 0.185 &  10^{-6} \\
2\times 10^{-3} &  10^{-6} & -0.05
\end{array}\),&&
\frac{g^{(2)}_+}{g} =\(\begin{array}{ccc}
0.140 &  10^{-3} &  10^{-3} \\
 10^{-3} & 0.139 &  10^{-3} \\
 10^{-3} &  10^{-3} & -0.05
\end{array}\);\nonumber\\
\frac{g^{(1)}_-}{g} =\(\begin{array}{ccc}
0.188 & 2\times 10^{-6} & 4\times10^{-4} \\
4\times10^{-4} & 0.183 &  10^{-3} \\
2\times 10^{-3} &   10^{-3} & -0.17
\end{array}\),&&
\frac{g^{(2)}_-}{g} =\(\begin{array}{ccc}
0.140 &  10^{-6} & 3\times  10^{-4} \\
  10^{-6} & 0.137 & 2\times  10^{-3} \\
3\times 10^{-4} & 2\times  10^{-3} & -0.140
\end{array}\).\nonumber\\
\end{eqnarray}
Constraints from top quark physics are significantly weaker, so this is not expected to affect the 
bound on $\mu$.

\section{Summary}\label{summary}
We have presented a variety of tools useful for studying fermion physics in soft-wall backgrounds, focusing heavily on the treatment of fermion masses. The equations of motion are non-trivial to solve and generically require numerical techniques. However, we have documented several special cases for which it is possible to decouple the equations of motion. These cases serve as useful examples for qualitatively understanding the full parameter space behavior, as they illuminate independent ``axes'' of the parameter space along which fermion behavior can be understood in detail. The utility of our approach is due not only to the fact that it effectively reduces the problem to solving a set of one-dimensional Schr\"{o}dinger-like equations, for which many theoretical and numerical tools have been created, but also to the fact that it applies to {\it any} soft-wall model in AdS space. This opens up the possibility of analyzing fermions in a wide variety of Higgs models. For example, it should be possible to analyze fermion physics in unHiggs scenarios, such as that considered in Ref.\cite{Falkowski:2008yr}, or to study other power-law Higgs behavior, as has been examined in the degenerate case in Refs.\cite{Batell:2008me,Aybat:2009mk}.

Furthermore, we have outlined methods for calculating fermion masses and wavefunctions in an arbitrary background with multiple flavors and arbitrary bulk parameters. The formalism maintains the orthogonality of the KK tower, making it particularly useful for studying the experimental consequences of soft-wall models with additional bulk fields. For example, we showed explicitly how to calculate the fermionic couplings to KK gauge bosons. The off-diagonal entries in the coupling matrix are directly related to the amplitudes of flavor changing neutral current processes. Moreover, we argued that the experimental constraints on new sources of flavor violation are generically tighter in soft-wall models than in hard-wall models due to the smaller spacing of the KK resonances. While the tightening is not too constraining in a model with a {\it quadratic} dilaton, for an IR cut-off growing much less quickly than $z^2$, the constraints can become severe for model building.

We described a very simple numerical technique for calculating fermion spectra, which we used to show the full parameter space dependence of SM fermion masses on the bulk mass parameters. The technique maintains the attractive features of the formalism, such as an orthogonal KK tower and extends naturally to incorporate several generations of fermions. Thus, the solutions allow for the straightforward calculation and interpretation of new physics observables.

Both the analytical and numerical results suggest the potential for rich collider physics that is substantially different from that obtained in hard-wall models. While our results were for a particular choice of a bulk Higgs VEV, they demonstrate that a soft-wall background can lead to a distinctive phenomenology. For example, our results indicate the presence of a lower bound on fermion masses in a large area of the parameter space, suggestive of a seesaw-like mechanism. This could easily be implemented to explain neutrino masses in a scenario where the hierarchy between $k$ and $\mu$ is of order the GUT scale. By introducing three additional right handed neutrino fields, a fairly random difference between the bulk masses of the right-handed neutrinos and charged leptons could naturally lead to light neutrino masses. This would be an interesting variation on the ideas that are well-known in the hard-wall picture (cf. \cite{Grossman:1999ra,Huber:2000ie,Agashe:2003zs}).

Even with these differences, the essential and attractive features of the hard-wall can be retained in our model. We presented results for three generations of fermions with anarchic 5D parameters that reveal standard model-like particle masses and GIM-like suppression of KK gauge boson mediated flavor changing neutral currents. We argued that this implied a fairly modest bound on the KK scale. A more general analysis of flavor physics bounds will lead to stringent constraints in the soft-wall model. Having developed the tools needed to examine electroweak and flavor physics in full detail, a more detailed study can be now be undertaken.

\section*{Acknowledgements}
We thank Brian Batell, Thomas Kelley, Arkady Vainshtein, and Mikhail Voloshin for helpful discussions. The work of T.G. is supported by the Australian Research Council while that of D.S. is supported by a Fellowship from the School of Physics and Astronomy at the University of Minnesota.

\end{document}